# Superconductivity of centrosymmetric and non-centrosymmetric phases in antiperovskite (Ca,Sr)Pd$_3$P

Akira Iyo,[1,*] Izumi Hase,[1] Hiroshi Fujihisa,[1] Yoshito Gotoh,[1] Nao Takeshita,[1] Shigeyuki Ishida,[1] Hiroki Ninomiya,[1] Yoshiyuki Yoshida,[1] Hiroshi Eisaki[1], Hishiro T. Hirose,[2] Taichi Terashima,[2] Kenji Kawashima,[1,3]

[1]National Institute of Advanced Industrial Science and Technology (AIST), 1-1-1 Umezono, Tsukuba, Ibaraki 305-8568, Japan

[2]National Institute for Materials Science, Tsukuba, Ibaraki 305-0003, Japan

[3]IMRA JAPAN Co., Ltd., 2-36 Hachiken-cho, Kariya-shi, Aichi 448-8650, Japan

**Abstract**

In the recently discovered antiperovskite phosphide (Ca,Sr)Pd$_3$P, centrosymmetric (CS) and non-centrosymmetric (NCS) superconducting phases appear depending on the Sr concentration, and their transition temperatures ($T_c$) differ by as much as one order of magnitude. In this study, we investigated the superconducting properties and electronic band structures of CS orthorhombic (CS$o$) (Ca$_{0.6}$Sr$_{0.4}$)Pd$_3$P ($T_c$ = 3.5 K) and NCS tetragonal (NCS$t$) (Ca$_{0.25}$Sr$_{0.75}$)Pd$_3$P ($T_c$ = 0.32 K) samples with a focus on explaining their large $T_c$ difference. Specific heat measurements indicated that CS$o$ (Ca$_{0.6}$Sr$_{0.4}$)Pd$_3$P was an s-wave superconductor in a moderate-coupling regime with a 2$\Delta_0$/k$_B$$T_c$ value of 4.0. Low-lying phonons leading to the strong coupling in the structurally analogous SrPt$_3$P were unlikely to be present in CS$o$ (Ca$_{0.6}$Sr$_{0.4}$)Pd$_3$P. Given that CS$o$ (Ca$_{0.6}$Sr$_{0.4}$)Pd$_3$P and NCS$t$ (Ca$_{0.25}$Sr$_{0.75}$)Pd$_3$P exhibited similar Debye temperatures ($\Theta_D$) of approximately 200 K, the large $T_c$ difference could not be attributed to $\Theta_D$. $T_c$ of each phase was accurately reproduced based on the Bardeen–Cooper–Schrieffer (BCS) theory using experimental data and the density of states of the Fermi level $N(0)$ calculated from their band structures. We concluded that the considerable suppression of $T_c$ in NCS$t$ (Ca$_{0.25}$Sr$_{0.75}$)Pd$_3$P can be primarily attributed to the decrease in $N(0)$ associated with the structural phase transition without considering the lack of inversion symmetry.

Key words: antiperovskite, superconductor, non-centrosymmetric space group, superconducting property, electronic band structure

*Corresponding author.
E-mail address: iyo-akira@aist.go.jp (Akira Iyo)

## 1. Introduction

In recent years, NCS superconductors whose crystal structures lack inversion symmetry have attracted considerable attention for their potential to realize unconventional superconductivity. The search for new materials and the measurements of superconducting properties have been actively conducted for NCS superconductors [1,2]. Recently, we successfully synthesized antiperovskite phosphides $M$Pd$_3$P ($M$ = Ca, Sr, Ba) [3] in the search for Mg$_2$Rh$_3$P-related materials [4]. In addition, we discovered a hidden superconducting phase in the solid solution between CaPd$_3$P and SrPd$_3$P, namely, (Ca$_{1-x}$Sr$_x$)Pd$_3$P [3]. (Ca$_{1-x}$Sr$_x$)Pd$_3$P has a unique phase diagram owing to the diversity of its crystal structure and superconductivity and crystal structure as follows. The endmember CaPd$_3$P exhibited a structural phase transition from CS$o$ (*Pnma*) to NCS orthorhombic

(NCS$o$) (*Aba*2) near room temperature (RT) when the temperature was lowered. The phase transition temperature was decreased by a Sr substitution for Ca, and superconductivity was induced in the CS$o$ phase (0.17 ≤ $x$ ≤ 0.55) at a critical temperature ($T_c$) of approximately 3.5 K. For further Sr substitution (0.6 ≤ $x$ ≤ 1.0), the NCS$t$ (*I*4$_1$*md*) phase was stabilized, and the superconductivity was suppressed by more than an order of magnitude: $T_c$ = 0.32 K and 0.06 K for $x$ = 0.75 and 1.0, respectively.

The antiperovskite (Ca$_{1-x}$Sr$_x$)Pd$_3$P has a three-dimensional network of corner-sharing Pd$_6$P octahedra, and the structural phase transitions are caused by the rotation and deformation of the Pd$_6$P octahedra and displacement of the P atoms inside the octahedra. Here, a simple question arises as to why such a large difference in $T_c$ occurs between the CS$o$ and NCS$t$ phases, even though the basic structure remains the same. According to the BCS equation, $T_c$ depends on $\Theta_D$, $N(0)$, and the effective attractive potential ($V$). In addition, in the case of NCS superconductors, the lack of inversion symmetry can decrease $T_c$. In general, the hybridization of spin-singlet and spin-triplet pairs can be present in compounds without inversion symmetry. Superconductivity is disadvantageous for such hybridization. For example, the reason for the lower $T_c$ (= 2.8 K) of Li$_2$Pt$_3$B than that (= 8 K) of Li$_2$Pd$_3$B is considered to be due to the hybridization of spin-triplet pairs [5-7]. Then, another question arises when the CS$o$ phase in (Ca,Sr)Pd$_3$P is compared with the analogous antiperovskites SrPt$_3$P and CaPt$_3$P. SrPt$_3$P and CaPt$_3$P have also CS structures (*P*4/*nmm*) but show strong-coupling superconductivity with higher $T_c$ of 8.4 K and 6.6 K, respectively [8]. Thus, why does the CS$o$ phase exhibit only about half the $T_c$ of SrPt$_3$P and CaPt$_3$P, despite their structure similarity?

The objective of this study is to clarify the physical properties of the two superconducting phases in (Ca$_{1-x}$Sr$_x$)Pd$_3$P and provide explanations for the above questions. For this purpose, CS$o$ (Ca$_{0.6}$Sr$_{0.4}$)Pd$_3$P and NCS$t$ (Ca$_{0.25}$Sr$_{0.75}$)Pd$_3$P were selected as representative samples because they have approximately central compositions in the CS$o$ and NCS$t$ phases in (Ca,Sr)Pd$_3$P, respectively. First, we show structure refinements of the samples and discuss the effect of the Sr substitution for Ca on the structural phase transition by the analysis of systematic structural data. Next, we indicate measurements of magnetization, resistivity, specific heat and pressure effect for the samples and derive superconducting parameters. Finally, we show electronic band structures of the CS and NCS phases and discuss the cause of the large $T_c$ difference using the theoretically and experimentally obtained parameters.

## 2. Materials and method

CS$o$ (Ca$_{0.6}$Sr$_{0.4}$)Pd$_3$P and NCS$t$ (Ca$_{0.25}$Sr$_{0.75}$)Pd$_3$P polycrystalline samples were synthesised through a solid-state reaction by using Pd powder and precursors with nominal compositions of CaP and SrP. A sample with a nominal composition of (Ca$_{0.6}$Sr$_{0.4}$)Pd$_3$P or (Ca$_{0.25}$Sr$_{0.75}$)Pd$_3$P was ground using a mortar in an N$_2$-filled glove box. The ground powder was pressed into a pellet and subsequently enclosed in an evacuated quartz tube. The sample was heated at 970 °C for 12 h, followed by furnace cooling. The details of the sample preparation are described elsewhere [3]. The composition ratios of the samples were measured using an energy-dispersive X-ray spectrometer (Oxford, SwiftED3000) equipped with an electron microscope (Hitachi High-Technologies, TM3000).

Powder X-ray diffraction patterns were obtained at RT (~ 295 K) using a diffractometer (Rigaku, Ultima IV) with Cu*Kα* radiation. Crystal structures were refined via Rietveld analysis using BIOVIA Materials Studio Reflex software (version 2018 R2) [9]. Magnetization ($M$) measurements were performed under various magnetic fields ($H$) using a magnetic-property measurement system (Quantum Design, MPMS-XL). The electrical resistivity ($\rho$) and specific heat ($C$) were measured using a physical property measurement system (Quantum Design, PPMS). The electrical resistivity of NCS*t* (Ca$_{0.25}$Sr$_{0.75}$)Pd$_3$P below 2 K was measured by cooling the sample with a dilution refrigerator as parameters of temperature and magnetic field. The temperature dependence of magnetization was measured under pressures (*P*) of up to 0.8 GPa using a piston–cylinder pressure cell. Daphne Oil 7474 was used as the pressure-transmitting medium [10]. The applied pressure was determined using a Pb manometer placed adjacent to the sample. The magnetization and specific heat were not measured for NCS*t* (Ca$_{0.25}$Sr$_{0.75}$)Pd$_3$P, given that its $T_c$ (~0.32 K) was lower than the measurable range of the apparatus.

First-principles electronic band structure calculations were performed using the full-potential linearized augmented plane-wave method, and a generalized gradient approximation was employed for the exchange-correlation potential [11]. To avoid complex calculations caused by the mixing of Sr and Ca at the same crystallographic site, we constructed three simplified models, namely CaPd$_3$P_CS*o* (constructed using the crystal structure parameters listed in Table 1), CaPd$_3$P_NCS*t*, and SrPd$_3$P_NCS*t* (both constructed using the crystal structure parameters listed in Table 2). By comparing CaPd$_3$P_CS*o* and CaPd$_3$P_NCS*t*, we can extract changes in the electronic structure due to differences in the crystal structure, and by comparing CaPd$_3$P_NCS*t* and SrPd$_3$P_NCS*t*, we can extract changes in the electronic structure due to differences in elements (Ca or Sr). For self-consistent loop calculations, a 250 *k*-point mesh for CaPd$_3$P_CS*o* (*I*4$_1$*md*, *Z* = 16) and a 1000 *k*-point mesh for CaPd$_3$P_NCS*t* and SrPd$_3$P_NCS*t* (*Pnma*, *Z* = 4) were used. For the density of states (DOS) calculations, we used 2000 *k*-points for all three models. The entire calculation was implemented using the WIEN2k computer program [12]. The details of our calculations are presented elsewhere [13].

## 3. Results and discussion

### 3.1 Crystal structure

Figures 1 (a) and (b) present the Rietveld fittings for the diffraction patterns of CS*o* (Ca$_{0.6}$Sr$_{0.4}$)Pd$_3$P and NCS*t* (Ca$_{0.25}$Sr$_{0.75}$)Pd$_3$P at RT, respectively. Composition ratios were measured as Ca:Sr:Pd:P = 0.57(1):0.39(1):3.04(2):1.01(1) for CS*o* (Ca$_{0.6}$Sr$_{0.4}$)Pd$_3$P and 0.24(1):0.74(1):3.02(3):1.00(2) for NCS*t* (Ca$_{0.25}$Sr$_{0.75}$)Pd$_3$P, which were in good agreement with the nominal compositions. The Rietveld fittings were performed by fixing the occupancy of each site to one and setting the ratio of Sr and Ca in the virtual chemical species to the same values as the nominal composition ratios.

Their diffraction patterns were well-fitted with a weighted-profile reliability factor ($R_{wp}$) of approximately 10%. Note that $R_{wp}$ values were obtained from the Rietveld fittings to the diffraction patterns with background subtracted beforehand. Refined structure parameters of CS*o* (Ca$_{0.6}$Sr$_{0.4}$)Pd$_3$P and NCS*t* (Ca$_{0.25}$Sr$_{0.75}$)Pd$_3$P are summarized in Table 1 and 2, respectively. Schematics of the refined crystal structures are shown in the insets. The structures of CS*o* (Ca$_{0.6}$Sr$_{0.4}$)Pd$_3$P and NCS*t* (Ca$_{0.25}$Sr$_{0.75}$)Pd$_3$P are analogous to those of the CS strong-coupling superconductor *A*Pt$_3$P (*A* = La, Ca, Sr) [8] and the heavy fermion NCS superconductor CePt$_3$Si ($T_c$ = 0.75 K) [14], respectively. As can be seen in the inset of Fig. 1(b), the lack of inversion symmetry in NCS*t* (Ca$_{0.25}$Sr$_{0.75}$)Pd$_3$P is due to the shift of P atoms along the *c*-axis direction.

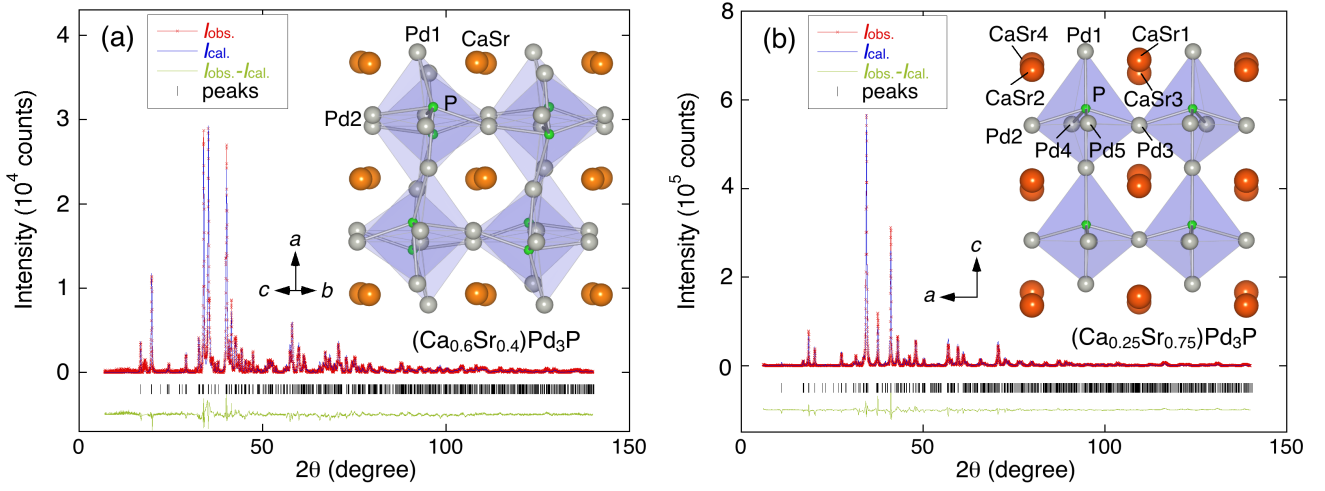

Fig. 1. Rietveld fittings for (a) CS$o$ (Ca$_{0.6}$Sr$_{0.4}$)Pd$_3$P and (b) NCS$t$ (Ca$_{0.25}$Sr$_{0.75}$)Pd$_3$P. $I_{obs.}$ and $I_{cal.}$ indicate the observed and calculated diffraction intensities, respectively. Refined crystal structures illustrated in the insets were obtained using VESTA software [15].

Table 1. Structure parameters obtained by the Rietveld structure refinement for CS$o$ (Ca$_{0.6}$Sr$_{0.4}$)Pd$_3$P at RT.

| Atoms | WP | x | y | z |
|---|---|---|---|---|
| CaSr | 4c | −0.0035(6) | 1/4 | 0.7805(6) |
| Pd$_1$ | 4c | 0.4543(3) | 1/4 | 0.2042(4) |
| Pd$_2$ | 8d | 0.2268(2) | 0.0212(3) | 0.4570(2) |
| P | 4c | 0.1992(7) | 1/4 | 0.1789(9) |

$Pnma$ (orthorhombic no. 62), $a$ = 9.0022(3) Å, $b$ = 6.1732(2) Å, $c$ = 6.5112(2) Å, $V$ = 361.85 Å$^3$ ($Z$ = 4), $R_{wp}$ = 10.45%, $U_{iso}$ = 0.028(1) for all atoms. WP: Wyckoff positions.

Table 2. Structure parameters obtained by the Rietveld structure refinement for NCS$t$ (Ca$_{0.25}$Sr$_{0.75}$)Pd$_3$P at RT.

| Atoms | WP | x | y | z |
|---|---|---|---|---|
| CaSr$_1$ | 4a | 0 | 0 | 0.0101 |
| CaSr$_2$ | 4a | 0 | 1/2 | −0.0112(13) |
| CaSr$_3$ | 4a | 1/2 | 0 | −0.0182(15) |
| CaSr$_4$ | 4a | 1/2 | 1/2 | −0.0003(10) |
| Pd$_1$ | 16c | 0.2480(6) | 0.2512(5) | 0.0280(8) |
| Pd$_2$ | 8b | 1/2 | 0.2626(8) | 0.1192(9) |
| Pd$_3$ | 8b | 0 | 0.2681(8) | 0.1194(9) |
| Pd$_4$ | 8b | 0 | 0.1856(7) | 0.3718(8) |
| Pd$_5$ | 8b | 1/2 | 0.2619(7) | 0.3735(9) |
| P | 16c | 0.2491(10) | 0.2526(10) | 0.1549(8) |

$I4_1md$ (tetragonal no. 109): $a$ = 8.7737(3) Å, $c$ = 19.1446(9) Å, $V$ = 1473.7(1) Å$^3$ ($Z$ = 16), and $R_{wp}$ = 10.11%. $U_{iso}$ = 0.045(1) Å$^2$ for all atoms. WP: Wyckoff positions.

The volume ($V_{oct}$) and bond angle variance ($\sigma^2$) of the Pd$_6$P octahedra calculated using the refined structural parameters were plotted for CS$o$ ($x$ = 0.25 and 0.4) and NCS$t$ ($x$ = 0.75 and 1.0) samples, as displayed in Fig. 2. Here, $\sigma^2$ is defined as $\sigma^2 = \frac{1}{11}\sum_{i=1}^{12}(\theta_i - 90)^2$, where $\theta_i$ is the Pd–P–Pd bond angle in the Pd$_6$P octahedra, which indicates the degree of distortion of the Pd$_6$P octahedra [16]. $V_{oct}$ increased with $x$ monotonically across both phases, which indicated that the substitution of the larger size element Sr for Ca acted as a negative chemical pressure on the $Pd_6P$ octahedra. The value of $\sigma^2$ increased in accordance with $x$ in each phase and considerably decreased at the phase boundary; which implies that the negative chemical pressure increased the distortion of the $Pd_6P$ octahedra and the structural phase transition occurred to avoid the energy increase associated with the distortion. Of note, $V_{oct}$ and $\sigma^2$ of the $Pt_6P$ octahedra in $CaPt_3P$ ($SrPt_3P$) were 14.6 (15.1) $Å^3$ and 771 (745) degree$^2$, respectively. The distortion of the $Pt_6P$ octahedra was much greater than that of the $Pd_6P$ octahedra. In contrast to $(Ca,Sr)Pd_3P$, the Sr substitution for Ca reduced the distortion of the $Pt_6P$ octahedra.

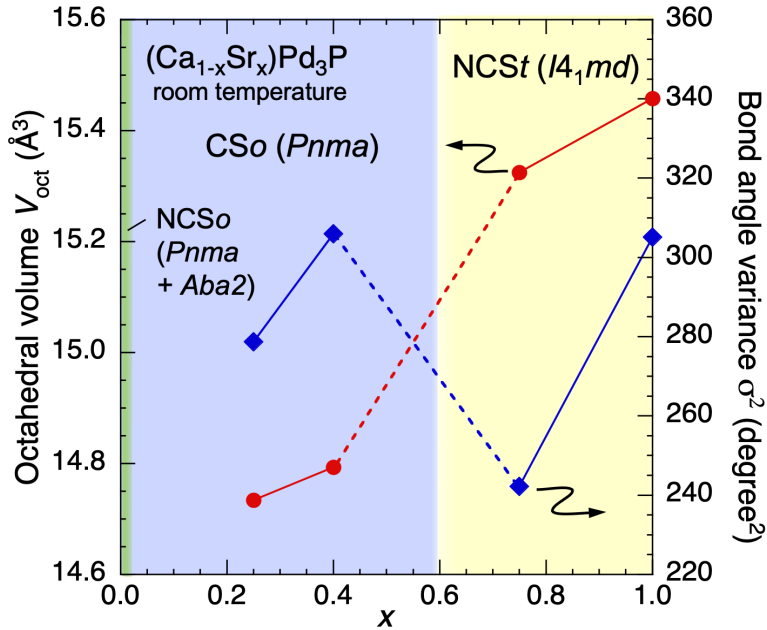

Fig. 2. Composition $x$ dependence of the volume $V_{oct}$ and bond angle variance $\sigma^2$ of $Pd_6P$ octahedra at RT for $x = 0.25$ and 0.4 (CS$o$ phase) and for $x = 0.75$ and 1.0 (NCS$t$ phase). The structural data in ref. [3] were used for $x = 0.25$ and 1.0.

3.2 Magnetization measurements

The temperature ($T$) dependence of $4\pi M/H$ for CS$o$ $(Ca_{0.6}Sr_{0.4})Pd_3P$ is shown in Fig. 3(a). A demagnetization correction was applied for $M$, considering the sample shape and size. The sample exhibited an abrupt diamagnetic transition at 3.38 K due to the occurrence of superconductivity, with a shielding volume fraction of approximately 100% ($4\pi M/H \sim -1$). Figure 3(b) presents the field dependence of the magnetization ($M$–$H$) curves for CS$o$ $(Ca_{0.6}Sr_{0.4})Pd_3P$ at various values of $T$ ($<T_c$). The magnetic field was applied parallel to the longitudinal direction of a rectangular sample with an aspect ratio of approximately four. The $M$–$H$ curves were that of a Type II superconductor. Each magnetization curve exhibited a broad minimum, and the minimum shifted to lower fields as $T$ increased. The lower critical fields $H_{c1}$ were evaluated from the $M$–$H$ curves as follows [17]: First, the demagnetization factor ($N$) of the sample was estimated from the initial slope (solid line in Fig. 3(b)) of the $M$–$H$ curve. The value of $N$ was then derived as 0.09 using $M = -H/4\pi(1-N)$. Then, $H_p$ for each curve was determined as the intersection point between the initial slope and the level line of the minimum magnetization, as indicated by the dashed line in Fig. 3(b). After that, $H_{c1}$ was calculated using $H_{c1} = H_p/(1 - N)$. As depicted in the inset of Fig. 3, $H_{c1}$ linearly changed for $(T/T_c)^2$, which was consistent with the Ginzburg–Landau (GL) theory, for which $H_{c1}(T) = H_{c1}(0)[1 - (T/T_c)^2]$. $H_{c1}$ at 0 K, denoted by $H_{c1}(0)$, was derived as 123(2) Oe by linear fitting to the data (see the inset of Fig. 3(b)).

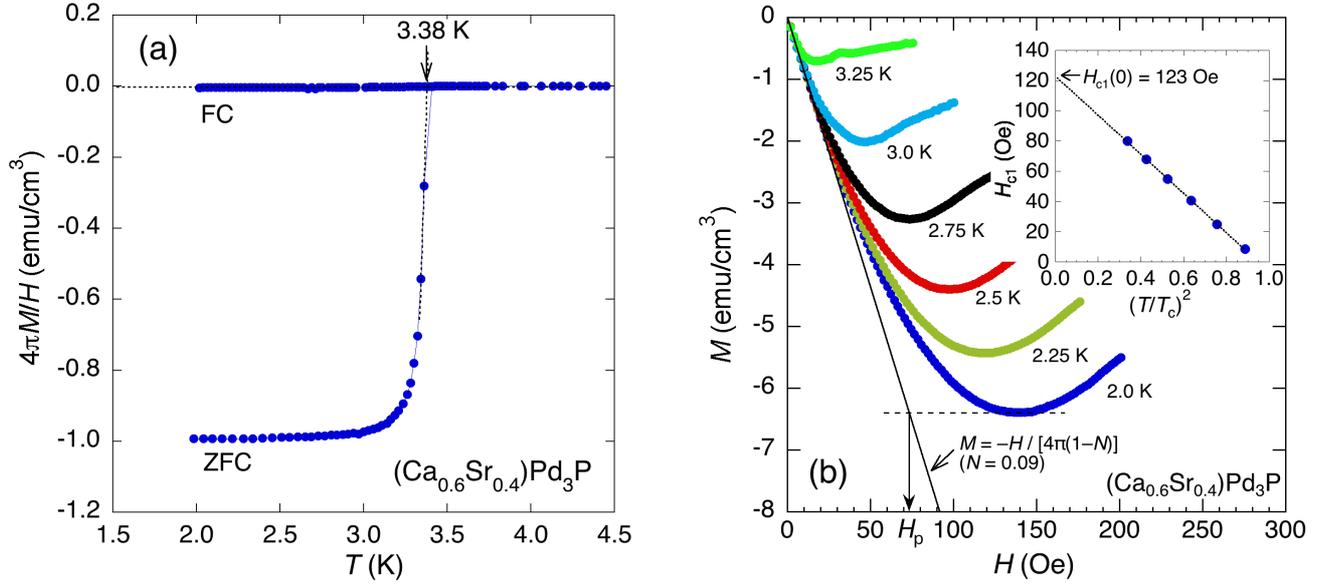

Fig. 3. (a) Temperature dependence of $4\pi M/H$ and (b) field dependence of magnetization ($M$–$H$) at various values of $T$ ($< T_c$) for CSo (Ca$_{0.6}$Sr$_{0.4}$)Pd$_3$P. The inset presents $H_{c1}$ with respect to $(T/T_c)^2$.

3.3 Resistivity measurements

Figure 4(a) presents the temperature dependence of $\rho$ for the CSo (Ca$_{0.6}$Sr$_{0.4}$)Pd$_3$P. The sample exhibited a metallic behavior in the normal state, with an *RRR* defined by a $\rho(300\ \mathrm{K})/\rho(5\ \mathrm{K})$ value of 1.79. A superconducting transition occurred at 3.58 K, with a transition width as small as 0.04 K. To approximate the Debye temperature $\Theta_\mathrm{D}$, we fitted $\rho(T)$ above 5 K using the parallel resistance model $1/\rho(T) = 1/\rho_\mathrm{BG}(T) + 1/\rho_\mathrm{max}$ [18], where $\rho_\mathrm{BG}(T)$ and $\rho_\mathrm{max}$ are the Bloch–Grüneisen term and saturation resistivity, respectively. A fit employing this model is indicated in Fig. 4(a) by the green curve, which reveals that $\Theta_\mathrm{D} = 178(3)$ K.

The temperature dependence of the resistivities under magnetic fields of up to 12 kOe is shown in Fig. 4(b) for CSo (Ca$_{0.6}$Sr$_{0.4}$)Pd$_3$P. The transition curves were shifted approximately parallel to lower temperature as the magnetic field increased, which indicated the excellent quality of the sample. The inset in Fig. 4(b) presents the $T$ dependence of the upper critical magnetic field $H_{c2}$, as evaluated from the midpoint of the resistive transitions. As observed, $H_{c2}(T)$ was not in accordance with the Werthamer–Helfand–Hohenberg (WHH) theory [19]. There were approximately linear changes in $H_{c2}$ over the measured temperature range. Therefore, $H_{c2}$ at 0 K, which is denoted by $H_{c2}(0)$, was determined as 25.3(1) kOe using linear extrapolation, as indicated by the dotted line in Fig. 4(b). The Ginzburg–Landau coherence length $\xi_0$ was calculated as 114(1) Å using $H_{c2}(0) = \Phi_0/2\pi\xi_0^2$, where $\Phi_0$ is the magnetic flux quantum. The London penetration depth $\lambda_0$ of $1.95(2) \times 10^3$ Å and the GL parameter $\kappa_\mathrm{GL}$ ($= \lambda_0/\xi_0$) of 17.1(1) were calculated using $H_{c1}(0) = \Phi_0 \ln\kappa_\mathrm{GL}/4\pi\lambda_0^2$. The $\kappa_\mathrm{GL}$ is a consistent value ($>1/\sqrt{2}$) for a Type II superconductor.

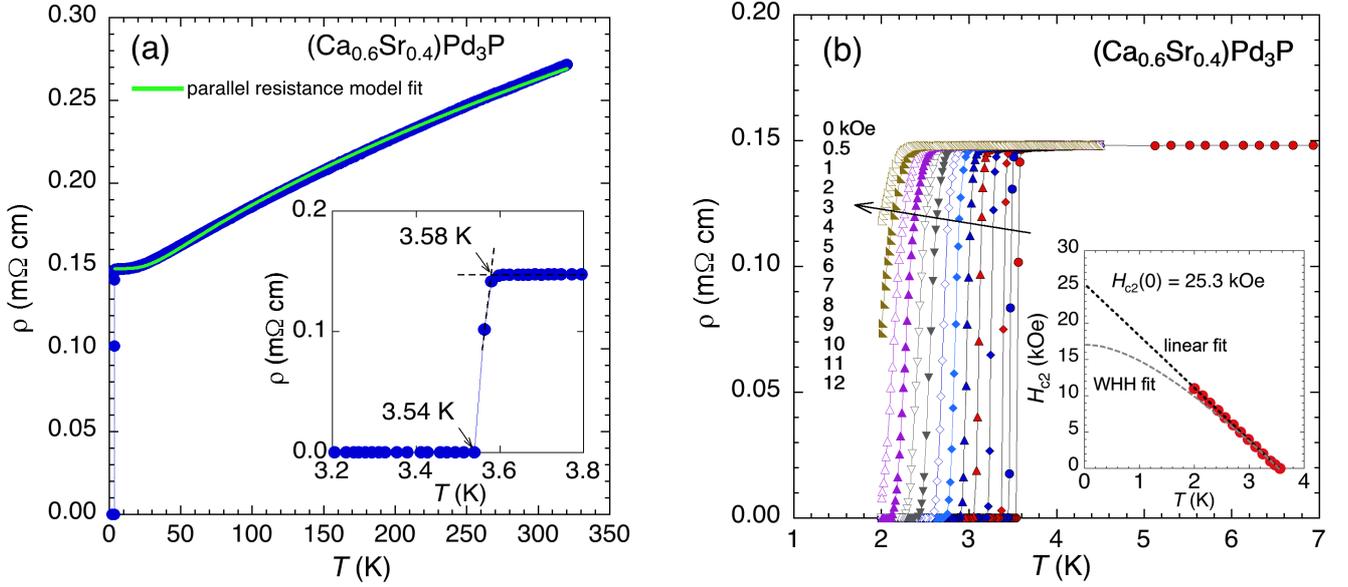

Fig. 4. (a) Temperature dependence of $\rho$ for CS$o$ (Ca$_{0.6}$Sr$_{0.4}$)Pd$_3$P. The inset presents a magnified view near $T_c$. The green curve indicates the fitting using the parallel resistance model. (b) Temperature dependence of $\rho$ around $T_c$ under magnetic fields of up to 12 kOe. The inset presents the temperature dependence of the upper critical magnetic field $H_{c2}$ evaluated from the midpoint of the resistive transitions. The linear fit and WHH fit (dirty limit) are indicated by a dotted line and dashed curve, respectively.

The temperature dependence of $\rho$ for NCS$t$ (Ca$_{0.25}$Sr$_{0.75}$)Pd$_3$P is illustrated in Fig. 5(a). The resistivity exhibited a similar temperature dependence as CS$o$ (Ca$_{0.6}$Sr$_{0.4}$)Pd$_3$P in the normal state, with an $RRR$ of 1.83. The superconducting transition was slightly broad, namely, an onset of transition at 0.70 K and zero resistivity at 0.32 K, as indicated in the inset of Fig. 5(a). The broadness of the superconducting transition could be attributed to the nonuniformity of composition $x$ in NCS$t$ (Ca$_{0.25}$Sr$_{0.75}$)Pd$_3$P. Such nonuniformity should also be present in CS$o$ (Ca$_{0.6}$Sr$_{0.4}$)Pd$_3$P ($x$ = 0.4); however, CS$o$ (Ca$_{0.6}$Sr$_{0.4}$)Pd$_3$P exhibited the remarkably narrow superconducting transition because of the little compositional dependence of $T_c$ around $x$ = 0.4 [3]. The zero-resistivity temperature was adopted as the $T_c$ of the bulk polycrystalline sample and used in the subsequent analysis. Using the same method as mentioned above, $\Theta_D$ was determined as 217(4) K for NCS$t$ (Ca$_{0.25}$Sr$_{0.75}$)Pd$_3$P. Unlike the case of CS$o$ (Ca$_{0.6}$Sr$_{0.4}$)Pd$_3$P, $H_{c2}$ of NCS$t$ (Ca$_{0.25}$Sr$_{0.75}$)Pd$_3$P was evaluated by measuring the field dependence of $\rho$ as a parameter of $T$ (= 0.04−1.70 K), as shown in Fig. 5(b). The inset in Fig. 5(b) presents the $T$ dependence of $H_{c2}$, where $H_{c2}$ is defined as the magnetic field at which $\rho$ become zero at each temperature. $H_{c2}(0)$ was determined as 0.32(1) kOe using linear extrapolation, as indicated by the dotted line in Fig. 5(b), and $\xi_0$ was calculated as 1.01(2) ×10$^3$ Å. $H_{c2}(0)$ was much smaller than the Pauli paramagnetic limiting field (~6 kOe), which can serve as a guide for the presence of spin-triplet pairing in NCS superconductors [1,2,14].

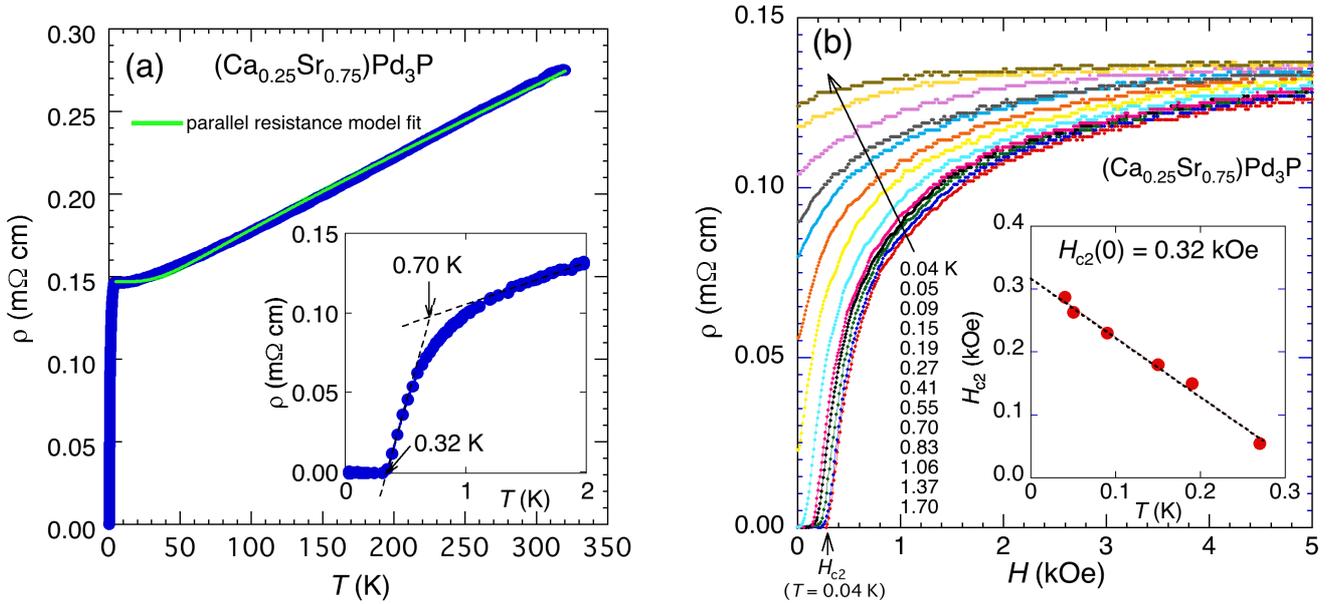

Fig. 5. (a) $T$ dependence of $\rho$ for NCS$t$ (Ca$_{0.25}$Sr$_{0.75}$)Pd$_3$P. The green curve presents the fitting using the parallel resistance model. The inset presents a magnified view near $T_c$. (b) $H$ dependence of $\rho$ at temperatures of 0.04–1.70 K. The inset presents the $T$ dependence of $H_{c2}$ defined as the magnetic field at which $\rho$ becomes zero at each temperature; an example of $H_{c2}$ determination at $T = 0.04$ K is illustrated in the figure. The linear fit is indicated by a dotted line.

3.4 Specific heat measurements

Figure 6(a) presents the $T^2$ dependence of $C/T$ in the magnetic fields $H = 0$ and 20 kOe for CS$o$ (Ca$_{0.6}$Sr$_{0.4}$)Pd$_3$P. An abrupt increase in the specific heat due to the occurrence of superconductivity was observed at 3.45 K ($T^2 = 11.9$ K$^2$), which was consistent with $T_c$ measured with respect to the magnetization and resistivity, thus demonstrating the bulk nature of superconductivity. The specific heat jump was completely suppressed by the application of a magnetic field of 20 kOe. $C/T$ below $T^2 = 50$ K$^2$ was fitted using $C/T = \gamma_n + \beta T^2 + \delta T^4$ (red dashed curve in Fig. 6(a)), where the Sommerfeld constants $\gamma_n$ and $\beta$ ($\delta$) are coefficients related to the electron and phonon contributions to the total specific heat $C$ ($= C_{el} + C_{ph}$), respectively. The fitting yielded $\gamma_n = 6.9(2)$ mJ mol$^{-1}$ K$^{-2}$, $\beta = 1.56(2)$ mJ mol$^{-1}$ K$^{-4}$ and $\delta = 2.2(4) \times 10^{-3}$ mJ mol$^{-1}$ K$^{-6}$. The Debye temperature $\Theta_D = 184(1)$ K was derived using the formula $\beta = N(12/5)\pi^4 R \Theta_D^{-3}$, where R = 8.314 J mol$^{-1}$ K$^{-1}$ and $N = 5$ (the number of atoms in the unit cell). $\Theta_D$ was in good agreement with the value (178 K) obtained by fitting the resistivity curve. According to the McMillan equation [20], the electron–phonon coupling constant $\lambda_{e-p}$ is given by $\lambda_{e-p} = (\mu^* \ln(1.45T_c/\Theta_D) - 1.04)/((1 - 0.62\mu^*)\ln(1.45T_c/\Theta_D) + 1.04)$, where $\mu^*$ is a Coulomb pseudopotential parameter. $\lambda_{e-p}$ was obtained as 0.66 for CS$o$ (Ca$_{0.6}$Sr$_{0.4}$)Pd$_3$P using $T_c = 3.45$ K, $\Theta_D = 184$ K, and the standard $\mu^*$ value (i.e., 0.13).

The inset in Fig. 6(a) presents the $T/T_c$ dependence of $C_{el}/\gamma_n T_c$. The electronic specific heat $C_{el}$ was obtained by subtracting the phonon part ($C_{ph} = \beta T^2 + \delta T^4$) from the total $C$. The normalized specific heat jump $\Delta C_{el}/\gamma_n T_c$ was estimated as 1.76, which is 23% larger than the value (~1.43) predicted by the BCS theory in the weak-coupling limit. To evaluate the deviation of the size of the superconducting gap at the zero temperature $\Delta_0$ from the BCS prediction, the $\alpha$-model was used to fit the specific heat below $T_c$ [21,22]. As indicated by the red curve in the inset, the data were reproduced well with $\alpha$ ($\Delta_0/k_B T_c$) = 2.0, where k$_B$ is the Boltzmann constant. $2\Delta_0/k_B T_c$, which is a good indicator of coupling strength, was 4.0. This value is 13% larger than the value (~3.53) predicted by BCS theory. Considering that typical strong-coupling superconductors have

$2\Delta_0/k_BT_c$ of approximately 4.5–5 [8,23-25], CS*o* (Ca$_{0.6}$Sr$_{0.4}$)Pd$_3$P can be considered an *s*-wave superconductor in a moderate-coupling regime.

CS*o* (Ca$_{0.6}$Sr$_{0.4}$)Pd$_3$P is in contrast to the strong-coupling superconductor SrPt$_3$P ($2\Delta_0/k_BT_c \sim 5$). Both are antiperovskites with CS structures. SrPt$_3$P exhibited almost the same $\Theta_D$ (~190 K) as CS*o* (Ca$_{0.6}$Sr$_{0.4}$)Pd$_3$P due to the similarity of crystal structure and constituent elements. However, the $T_c$ (~8.4 K) of SrPt$_3$P was more than double the value of that of CS*o* (Ca$_{0.6}$Sr$_{0.4}$)Pd$_3$P. The high $T_c$ in SrPt$_3$P can be attributed to the strong coupling of charge carriers with low-lying phonons [8,26,27]. The presence of the low-lying phonons in SrPt$_3$P is evidenced by the nonlinear behavior of $C/T$ with respect to $T^2$, in that $C/T$ deviates from the line extrapolated from the lower temperature data at $T^2$ of approximately 20 K$^2$. It is improbable for such low-lying phonons to be present in CS*o* (Ca$_{0.6}$Sr$_{0.4}$)Pd$_3$P, given that no clear nonlinearity was observed as illustrated in Fig. 6(a). The $\delta$ of CS*o* (Ca$_{0.6}$Sr$_{0.4}$)Pd$_3$P was approximately a quarter of that of SrPt$_3$P, which may have caused the difference between the coupling strengths of the analogous compounds of SrPt$_3$P and CS*o* (Ca$_{0.6}$Sr$_{0.4}$)Pd$_3$P. The characteristic low-lying phonons, such as rattling modes, also induce strong electron–phonon coupling, thus resulting in an enhanced $T_c$ in the $\beta$-pyrochlore oxide KOs$_2$O$_6$ ($T_c$ = 9.6 K) [24] and an intermetallic compound IrGe ($T_c$ = 4.7 K) [25].

Figure 6(b) presents the $T^2$ dependence of $C/T$ for NCS*t* (Ca$_{0.25}$Sr$_{0.75}$)Pd$_3$P in a zero magnetic field. The fitting of the $C/T$ below $T^2$ = 50 K$^2$ with $C/T = \gamma_n + \beta T^2 + \delta T^4$ (red dashed curve in Fig. 6(b)) yielded $\gamma_n$ = 3.61(9) mJ mol$^{-1}$ K$^{-2}$, $\beta$ = 0.69(1) mJ mol$^{-1}$ K$^{-4}$ and $\delta$ = 3.6(2) × 10$^{-3}$ mJ mol$^{-1}$ K$^{-6}$. The Debye temperature $\Theta_D$ was derived as 242(2) K, which is close to the value obtained by the analysis of the resistivity curve. $\lambda_{e-p}$ was 0.39 for NCS*t* (Ca$_{0.25}$Sr$_{0.75}$)Pd$_3$P as obtained by the McMillan equation using $T_c$ = 0.32 K, $\Theta_D$ = 242 K, and $\mu^*$ = 0.13. According to the McMillan equation, a higher-$T_c$ CS*o* (Ca$_{0.6}$Sr$_{0.4}$)Pd$_3$P is expected to have higher $\Theta_D$; however, there was no significant difference between the $\Theta_D$ values of NCS*t* (Ca$_{0.25}$Sr$_{0.75}$)Pd$_3$P and CS*o* (Ca$_{0.6}$Sr$_{0.4}$)Pd$_3$P. The lower-$T_c$ NCS*t* (Ca$_{0.25}$Sr$_{0.75}$)Pd$_3$P exhibited rather higher $\Theta_D$ than that of CS*o* (Ca$_{0.6}$Sr$_{0.4}$)Pd$_3$P. Thus, the difference between the $T_c$ values of the two phases can be attributed to factors other than $\Theta_D$.

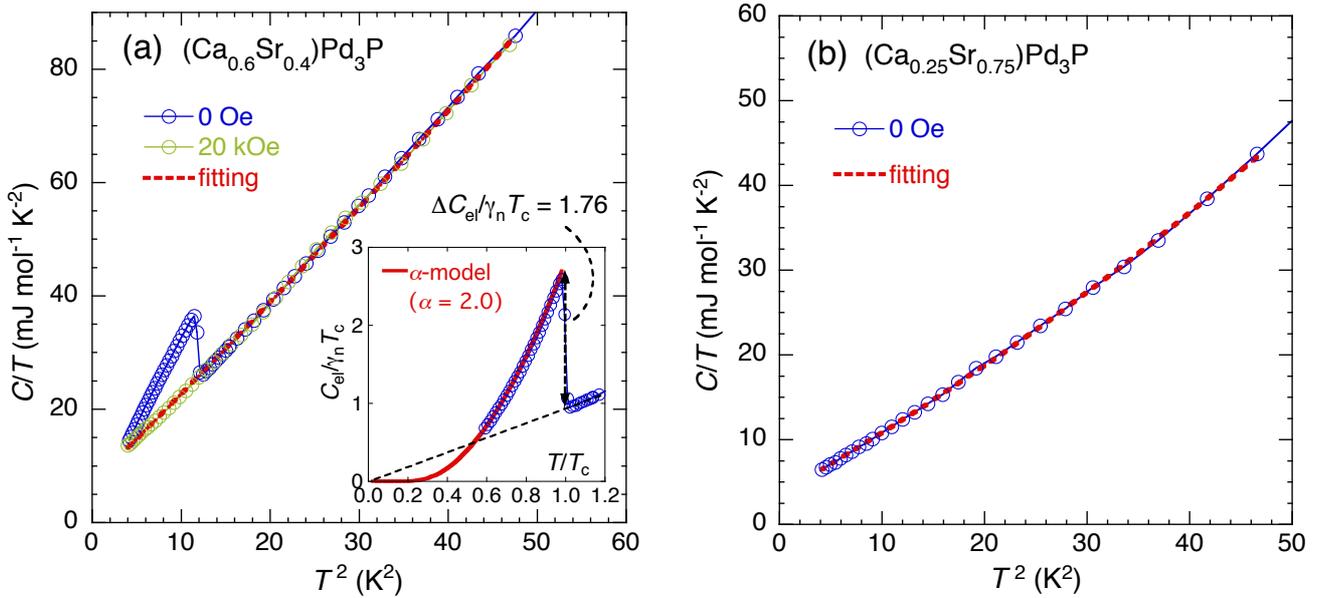

Fig. 6. $T^2$ dependence of $C/T$ for (a) CS*o* (Ca$_{0.6}$Sr$_{0.4}$)Pd$_3$P in magnetic fields $H$ = 0 and 20 kOe and (b) NCS*t* (Ca$_{0.25}$Sr$_{0.75}$)Pd$_3$P at $H$ = 0. The red dashed curves indicate the fitting of data for CS*o* (Ca$_{0.6}$Sr$_{0.4}$)Pd$_3$P ($H$ = 20 kOe) and NCS*t* (Ca$_{0.25}$Sr$_{0.75}$)Pd$_3$P ($H$ = 0 kOe) below $T^2$ = 50 K$^2$. The inset of Fig. 6(a) presents the $T/T_c$ dependence of $C_{el}/\gamma_n T_c$. The red solid curve in the inset represents a fitting using the $\alpha$-model.

3.5 Pressure effect measurements

Figure 7 presents the temperature dependence of the normalized magnetization of CS$o$ (Ca$_{0.6}$Sr$_{0.4}$)Pd$_3$P as a function of the applied pressure. As observed, $T_c$ decreased as the pressure increased, at a rate of d$T_c$/d$P$ = −0.15(1) K/GPa, as indicated in the inset of Fig. 7. The decrease in $T_c$ can be attributed to a decrease in the DOS at the Fermi level ($E_F$) $N(0)$ due to the compression of the lattice under pressure. The pressure effect of CS$o$ (Ca$_{0.6}$Sr$_{0.4}$)Pd$_3$P is in striking contrast with that of SrPt$_3$P. Jawdat et al. reported that $T_c$ of SrPt$_3$P increases from 8.35 K at the ambient pressure to a maximum of 8.49 K at a pressure of 0.90 GPa, with an initial slope of 0.2 K/ GPa [28]. Thereafter, $T_c$ decreases with the further application of pressure. The increase in $T_c$ in SrPt$_3$P is explained by an increase in the characteristic phonon energy. The increase in $T_c$ upon the application of pressure was also observed in β-pyrochlore oxides $A$Os$_2$O$_6$ ($A$ = Cs, Rb, K), where the low-energy phonons due to the rattling mode exist [29]. Thus, the contrasting pressure effect may indicate the absence of such characteristic phonons in CS$o$ (Ca$_{0.6}$Sr$_{0.4}$)Pd$_3$P.

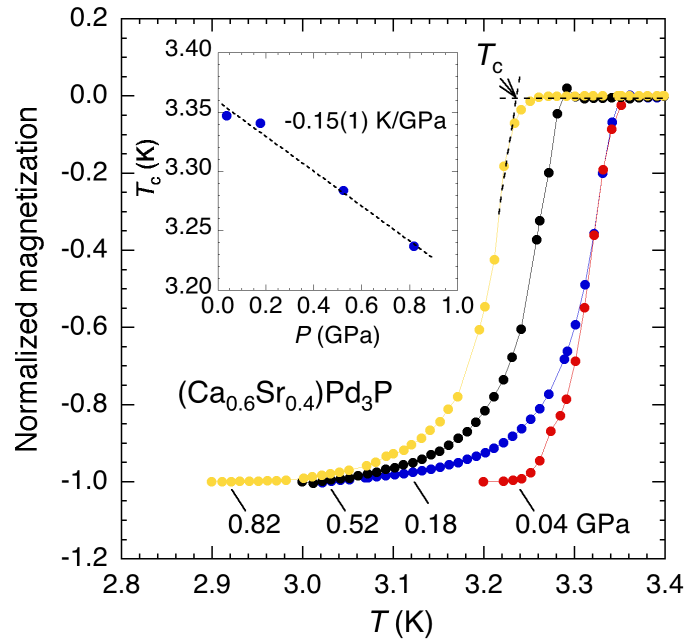

Fig. 7. Temperature dependence of the magnetization normalized to the normal-state and saturation diamagnetic magnetizations for CS$o$ (Ca$_{0.6}$Sr$_{0.4}$)Pd$_3$P. The inset presents the pressure dependence of $T_c$ determined using the method indicated in the main figure.

3.6 Band structure calculations

The DOSs of the three models, namely, CaPd$_3$P_CS$o$, CaPd$_3$P_NCS$t$, and SrPd$_3$P_NCS$t$, are shown in Fig. 8. Firstly, we point out that CaPd$_3$P_NCS$t$ and SrPd$_3$P_NCS$t$ exhibited almost identical DOS curves near $E_F$. This result is expected, given that the main contribution to $N(0)$ is from the Pd-$d$ orbitals. This behavior is very similar to that of SrPt$_3$P, which contains the Pt-$d$ orbitals as its main component [27,30]. Hence, as a first approximation, $A$-site cations (Ca or Sr) play a role in changing the crystal structure due to the different ionic radii. In contrast, the DOS curve near $E_F$ of CaPd$_3$P_CS$o$ was considerably different from those of CaPd$_3$P_NCS$t$ and SrPd$_3$P_NCS$t$. In CaPd$_3$P_CS$o$, $E_F$ was located at the peak of the DOS curve, whereas in CaPd$_3$P_NCS$t$ and SrPd$_3$P_NCS$t$, they were located in the dips of the DOS curves. As a result, CaPd$_3$P_CS$o$ exhibited a larger $N(0)$ than CaPd$_3$P_NCS$t$ and SrPd$_3$P_NCS$t$.

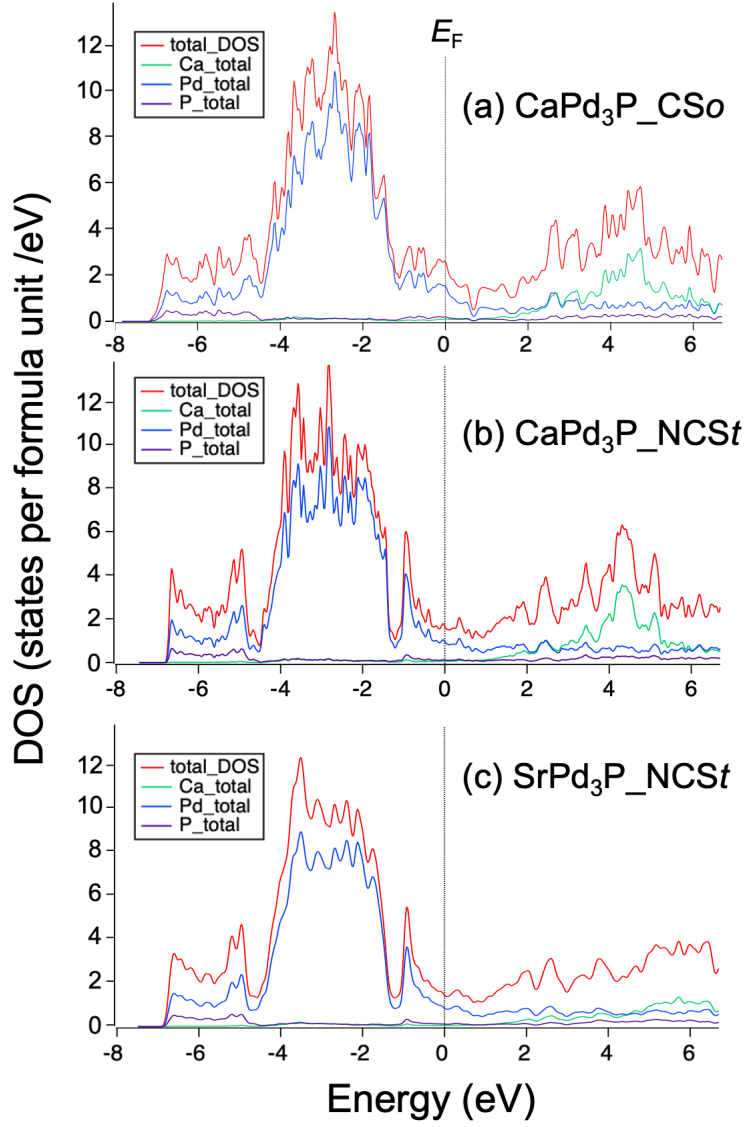

Fig. 8. DOSs of (a) CaPd$_3$P_CS$o$, (b) CaPd$_3$P_NCS$t$, and (c) SrPd$_3$P_NCS$t$, as obtained by the first-principles electronic band structure calculations.

The large $T_c$ difference between the two phases was discussed based on theoretical calculations. Given that superconductivity is a low-energy physics in general, $T_c$ is highly correlated with $N(0)$ in most of the theories of superconductivity, including the BCS and Migdal–Eliashberg theories. We theoretically found $N(0)$ per formula unit to be 2.65, 1.64, and 1.30 eV$^{-1}$, and the Sommerfeld constant $\gamma_{n\_calc}$ was calculated using the above $N(0)$ as 6.24, 3.88, and 3.06 for CaPd$_3$P_CS$o$, CaPd$_3$P_NCS$t$, and SrPd$_3$P_NCS$t$, respectively. Although the experimental $\gamma_n$ values of 6.94 and 3.61 mJ mol$^{-1}$ K$^{-2}$ for CS$o$ (Ca$_{0.6}$Sr$_{0.4}$)Pd$_3$P and NCS$t$ (Ca$_{0.25}$Sr$_{0.75}$)Pd$_3$P were smaller than the expected values $(1 + \lambda_{e\text{-}p})\gamma_{n\_calc}$ of 10.5 and 5.39 (or 4.25) mJ mol$^{-1}$ K$^{-2}$, respectively, they were in a relatively good agreement. For comparison, the values obtained from the calculations are summarized in Table 3, in addition to the experimental values.

Table 3. Parameters obtained for CS$o$ (Ca$_{0.6}$Sr$_{0.4}$)Pd$_3$P and NCS$t$ (Ca$_{0.25}$Sr$_{0.75}$)Pd$_3$P.

| Parameters | CS$o$ (Ca$_{0.6}$Sr$_{0.4}$)Pd$_3$P | NCS$t$ (Ca$_{0.25}$Sr$_{0.75}$)Pd$_3$P |
| --- | --- | --- |
| $T_c$ (K) | 3.54 ($\rho = 0$) | 0.32 ($\rho = 0$) |
| $H_{c1}(0)$ (Oe) | 123(2) | – |
| $H_{c2}(0)$ (kOe) | 25.3(1) | 0.32(1) |
| $\lambda_0$ (Å) | $1.95(2) \times 10^3$ | – |
| $\xi_0$ (Å) | 114(1) | $1.01(2) \times 10^3$ |
| $\kappa_{GL}$ | 17.1(1) | – |
| $\gamma_n$ (mJ mol$^{-1}$ K$^{-2}$) | 6.9(1) | 3.61(9) |
| $N(0)$ (eV$^{-1}$)[*1] | 2.65 | 1.64, 1.30 |
| $\gamma_{n\_cal}$ (mJ mol$^{-1}$ K$^{-2}$)[*1] | 6.24 | 3.88, 3.06 |
| $\beta$ (mJ mol$^{-1}$ K$^{-4}$) | 1.56(2) | 0.69(1) |
| $\Theta_D$ (K)[*2] | 184(1) | 242(2) |
| $\lambda_{e\text{-}p}$ | 0.66 | 0.39 |
| $\Delta C_{el}/\gamma_n T_c$ | 1.76 | – |
| $2\Delta_0/k_B T_c$ | 4.0 | – |
| $dT_c/dP$ (K/GPa) | –0.15(1) | – |

[*1] obtained by band structure calculations, [*2] obtained by specific heat measurements

As discussed in the previous section, the lower $T_c$ of NCS$t$ (Ca$_{0.25}$Sr$_{0.75}$)Pd$_3$P cannot be attributed to the changes in $\Theta_D$. According to the McMillan equation, the lower $T_c$ in NCS$t$ (Ca$_{0.25}$Sr$_{0.75}$)Pd$_3$P indicates a weaker electron–phonon coupling. However, the analysis of $T_c$ is not simple, given that $\lambda_{e\text{-}p}$ may depend on $N(0)$ and $\Theta_D$ in a complex manner based on the McMillan equation. Instead, we adopted a simpler BCS equation $T_c = 1.14\Theta_D \exp(-1/N(0)V)$. Inputting $T_c = 3.45$ K and $\Theta_D = 184$ K (experimental values for CS$o$ (Ca$_{0.6}$Sr$_{0.4}$)Pd$_3$P) and $N(0) = 2.65$ eV$^{-1}$ (calculated value for the model CaPd$_3$P_CS$o$) into this equation, we obtained $V = 0.0919$ eV for model CaPd$_3$P_CS$o$. We evaluated $T_c$ for model CaPd$_3$P_NCS$t$ using $N(0) = 1.64$ eV$^{-1}$ assuming that $V = 0.0919$ eV is also valid, and obtained $T_c = 0.36$ K. This $T_c$ value is in good agreement with the experimental value of 0.32 K for NCS$t$ (Ca$_{0.25}$Sr$_{0.75}$)Pd$_3$P. For model SrPd$_3$P_NCS$t$, $T_c = 0.064$ K was obtained using $N(0) = 1.30$ eV$^{-1}$, which is highly similar to the experimental value of 0.06 K for SrPd$_3$P [3]. Although the selection of the model requires further considerations, this $T_c$ estimation (i.e., 0.36–0.064 K) reproduced the experimental values accurately considering that the experimental sample was a mixed crystal of Sr and Ca. Thus, the considerable changes in the $T_c$ of CS$o$ (Ca$_{0.6}$Sr$_{0.4}$)Pd$_3$P and NCS$t$ (Ca$_{0.25}$Sr$_{0.75}$)Pd$_3$P can be primarily attributed to the considerable changes in $N(0)$.

Next, we discuss whether the absence of the inversion symmetry affects superconductivity in this system. In general, the spin-orbit interaction (SOI), which is mostly atomic interaction, causes an anti-symmetric spin-orbit coupling (ASOC) in the absence of the crystal inversion symmetry. The large SOI of Pt-5$d$ are speculated to have caused a large ASOC in NCS superconductors CePt$_3$Si and Li$_2$Pd$_3$B, and as a result, the odd-parity component of the superconducting order parameter is considerably mixed [31,32]. As shown above, the lower $T_c$ of the NCS$t$ (Ca$_{0.25}$Sr$_{0.75}$)Pd$_3$P can be attributed to its small $N(0)$. The absence of inversion symmetry typically decreases $T_c$ by mixing the odd-parity component of the order parameter. However, in NCS$t$ (Ca$_{0.25}$Sr$_{0.75}$)Pd$_3$P, the BCS theory (assuming no such mixing) reproduced the experimental $T_c$ well. This finding implies that the effect of such mixing is not significant in NCS$t$ (Ca$_{0.25}$Sr$_{0.75}$)Pd$_3$P, unlike NCS superconductors CePt$_3$Si and Li$_2$Pt$_3$B. Li$_2$Pd$_3$B represents a case in which the absence of inversion symmetry does not affect superconductivity. In Li$_2$Pd$_3$B and NCS$t$ (Ca$_{0.25}$Sr$_{0.75}$)Pd$_3$P, the most significant orbitals are the Pd-4$d$ orbitals which exhibit smaller SOI than the Pt-5$d$ orbitals in CePt$_3$Si and Li$_2$Pt$_3$B.

## 4. Conclusions

The measurements of the physical properties and calculations of the electronic band structure were performed for the CS and NCS phases in antiperovskite (Ca,Sr)Pd$_3$P. Systematic analysis of the structure parameters revealed that the substitution of Sr for Ca in (Ca,Sr)Pd$_3$P resulted in a negative chemical pressure on the Pd$_6$P octahedra and the structural phase transition occurred to reduce the distortion increased by the Sr substitution. CS$o$ (Ca$_{0.6}$Sr$_{0.4}$)Pd$_3$P was found to be an $s$-wave superconductor in a moderate-coupling regime, which is in contrast to the strong-coupling superconductor SrPt$_3$P. The weaker coupling in CS$o$ (Ca$_{0.6}$Sr$_{0.4}$)Pd$_3$P was possibly due to the absence of low-lying phonons. We found that the considerable differences in $T_c$ between CSo (Ca$_{0.6}$Sr$_{0.4}$)Pd$_3$P and NCSt CS$o$ (Ca$_{0.6}$Sr$_{0.4}$)Pd$_3$P was accurately explained based on the BCS theory using the experimentally obtained parameters and the theoretically calculated $N(0)$ based on the BCS theory, which implied that the mixing of the odd-parity component of the order parameter is not significant in NCS$t$ (Ca$_{0.25}$Sr$_{0.75}$)Pd$_3$P. Considering the contrasting superconductivity and structural nature of (Ca,Sr)Pd$_3$P and SrPt$_3$P, their solid solution (Ca$_{1-x}$Sr$_x$)(Pd$_{1-y}$Pt$_y$)$_3$P would be an intriguing platform in further studying the influence of the crystal structure on superconductivity.

## Acknowledgements

This work was supported by the JSPS KAKENHI (grant numbers JP19K04481, JP16H6439, JP19K03731, and JP19H05823) and the TIA collaborative research program KAKEHASHI "Tsukuba–Kashiwa–Hongo Superconductivity Kakehashi Project". We thank Dr. Takashi Yanagisawa and Dr. Yoichi Higashi for fruitful discussions.